\documentclass[showpacs,aps,prb,twocolumn]{revtex4}
\usepackage{amsmath}
\usepackage{amsfonts}
\usepackage{amssymb}
\usepackage{amsthm}
\usepackage{graphicx}

\begin{document}

\title{Current-induced synchronized magnetization reversal of two-body
Stoner particles with dipolar interaction}
\author{Z. Z. Sun}
\author{J. Schliemann}
\affiliation{Institute for Theoretical Physics, University of Regensburg,
D-93040 Regensburg, Germany}
\date{\today}

\begin{abstract}
We investigate magnetization reversal of two-body uniaxial Stoner particles,
by injecting spin-polarized current through a spin-valve
structure. The two-body Stoner particles perform synchronized dynamics and
can act as an information bit in computer technology.
In the presence of magnetic dipole-dipole interaction (DDI)
between the two particles, the critical switching current $I_c$ for
reversing the
two dipoles is analytically obtained and numerically verified in two typical
geometric configurations. $I_c$ bifurcates at a critical DDI strength, where
$I_c$ can be decreased to about $70\%$ of the usual value without DDI.
Moreover, we also numerically investigate the magnetic
hysteresis loop, magnetization self-precession, reversal time and the
synchronization stability phase diagram for the two-body system in the
synchronized dynamics regime.
\end{abstract}
\maketitle
\section{Introduction}

Recent advances on nanomagnetism technology allow to fabricate Stoner
particles (single-domain magnetic nanoparticles due to strong exchange
interaction)\cite{Hehn,Stamm,Shouheng,Zitoun}, which has attracted much
attention both from a fundamental point of view and potential
applications in information industry\cite{Book1,Book2}. Magnetization
dynamics of a single Stoner particle has been extensively studied, either
by static or time-dependent magnetic fields
\cite{Stoner,Doyle,Hiebert,Crawford,Acremann,Hillebrands,Back,Schumacher,Thirion,llg1,Sun1}, or by spin-polarized current
 approaches\cite{Tsoi,Slonczewski1,Zhang,Waintal,Wang3}. The current-induced
magnetization switching attracts much interest due to the locality and
convenient controllability of currents. However, the high critical switching
current $I_c$ (or density) hinders practical applications\cite{JSun}.
Many efforts were made for lowering the $I_c$, for instance, designing an
optimized spin current pattern\cite{Wang3}, using pure spin current\cite{Lu}
and thermal activation\cite{Hatami,Xia1}. Recently lower $I_c$ has also been
proposed in ferromagnetic semiconductors\cite{Nguyen,Garate}.

Moreover, since magnetic nanoparticles are actually fabricated in
arrays\cite{Hehn,Stamm,Shouheng,Zitoun},
the dipole-dipole interaction (DDI) between them will be important for the
magnetic switching behavior. Indeed there were several theoretical studies
on systems of two Stoner particles, the simplest case to investigate the
DDI effect, concluding the dipolar interaction can assist
the magnetic switching\cite{Bertram,Chen,Lyberatos,Huang,Xu,Pham,Plamada}.
Recently a collaboration including the present authors revisited the
two-body Stoner particles problem and proposed
a novel technological perspective by showing that,
in a synchronized dynamics mode
(i.e. both the magnetic vectors of the particles runs synchronously), the
critical switching field in presence of magnetic DDI can be dramatically
lowered. This implies even zero-field switching
to be achieved by appropriately engineering the DDI strength by adjusting the
interparticle distance\cite{Sunddi}. In the present paper,
we extend these studies by investigating
current-induced synchronized magnetization switching of such the
two-body Stoner particles system on injecting a spin-polarized current
through a spin-valve-like structure. As a result, the critical
switching current $I_c$ (or its density) can be lowered to about $70\%$ of the
usual value without DDI around a critical DDI strength.
This DDI strength corresponds to a distinguished
interparticle distance and has the same value as
in the previously studied case of field-induced switching\cite{Sunddi}. Furthermore, $I_c$ bifurcates at the critical DDI
strength by a square-root-like behavior on the DDI parameter. Moreover, we have
also numerically investigated issues including
the magnetic hysteresis
loop, magnetization self-precession, reversal time, and
the synchronization stability phase diagram for the two-body system in
presence of the DDI. These investigations revealed
many novel promising phenomena regarding future
device applications. This paper is organized as follows: In Sec. II the two
typical geometrical configurations of the two-body system in a spin-valve-like
setup are introduced along with the equations governing
current-induced magnetization dynamics.
Our analytical and numerical results are presented in
Sec. III and IV, respectively. We close with a discussion and concluding remarks
in Sec. V.

\section{Model}

\begin{figure}[htbp]
 \begin{center}
\includegraphics[width=7.5cm, height=4.cm]{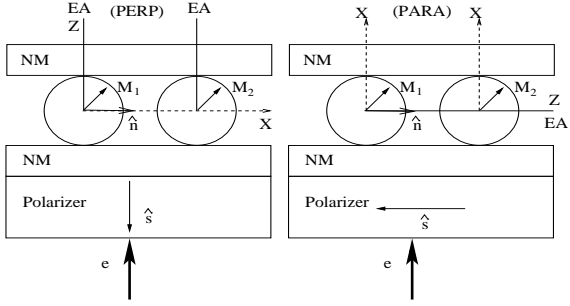}
 \end{center}
\caption{\label{fig1} A schematic diagram of two typical geometric
configurations (PERP and PARA) in spin-valve structures
and the corresponding coordinate axis orientations. Polarizer is a thick pinned ferromagnetic layer where $\hat{s}$ denotes its polarized direction and NM denotes nonmagnetic metallic layer. $EA$ denotes the easy axis and $e$ denotes the electrons flow. }
\end{figure}

We consider the spin-valve structure schematically shown in Fig.~\ref{fig1}. For simplicity, we consider two particles are in the same sphere forms. Two typical geometric configurations of the two-body system are investigated, where the unit vector $\hat{n}$ along the line connecting the two particles is either perpendicular or parallel to the magnetic easy axes (EA), assumed along $z$-axis, referred to as {\it PERP} configuration and {\it PARA} configuration, respectively. That is, if we define $\hat{n}=(\sin\theta_n\cos\phi_n, \sin\theta_n\sin\phi_n, \cos\theta_n)$ in the usual spherical coordinates where $z$-axis is the pole axis, thus $\theta_n=\pi/2$ for PERP configuration and $\theta_n=0,\pi$ for PARA (the two values are equivalent since DDI is invariant under $\hat{n} \mapsto -\hat{n}$). Moreover, without loss of generality, one can let $\phi_n=0$ for both configurations, i.e., $\hat{n}=\hat{x}$ in PERP and $\hat{n}=\hat{z}$ in PARA (see Fig.~\ref{fig1}).

The magnetization dynamics of two-body Stoner particles under a spin-polarized current is governed by the modified Landau-Lifshitz-Gilbert (LLG) equation\cite{Landau} with the spin-transfer torque (STT) term\cite{Slonczewski1,Zhang},
\begin{equation}
\dot{\vec{m}}_i= -\vec{m}_i\times\vec{h}^{t}_{i}
+\alpha \vec{m}_i\times \dot{\vec{m}}_i +\vec{T}_{stt}.\label{LLG1}
\end{equation}
Here $\vec{m}_i=\vec{M}_i/M_s$ is the normalized magnetization vector of the $i$th-particle $(i=1,2)$, $M_s=|\vec{M}_i|$ is the magnetization saturation of either particle, and $\alpha$ is the phenomenological Gilbert damping coefficient. We are assuming the two particles are completely identical with all the same concerned parameters for simplicity. The unit of time is set to be $(|\gamma|M_s)^{-1}$ where $\gamma$ is the gyromagnetic ratio. The total effective field $\vec{h}^{t}_i$ on each particle comes from the variational derivative of the total magnetic energy with respect to magnetization,
$\vec{h}^{t}_i=-\delta E/ \delta \vec{m}_i$. For the concerned system, the energy $E$ per particle volume $V$ (in units of $\mu_0M_s^2$; $\mu_0$: vacuum permeability) can be expressed as,
\begin{equation}
E=-\sum_{i=1,2} k m_{i,z}^2
+\eta[\vec{m}_1 \cdot\vec{m}_2-3(\vec{m}_1\cdot\hat{n})
(\vec{m}_2\cdot \hat{n})].\label{energy}
\end{equation}
Here the uniaxial magnetic anisotropic parameter $k$ summarizes both shape and magnetocrystalline contributions to the magnetic anisotropy along the easy axis and we only consider no external magnetic field case. The parameter $\eta\equiv \frac{V}{4\pi d^3}$ is
a geometric factor characterizing the DDI strength, where $d$ is the fixed distance between the two particles. We also omit the exchange interaction energy between two particles as the same reasons discussed in Ref.~\onlinecite{Sunddi}. The LLG equation can be written in the usual spherical coordinates, by putting $\vec{m}_i=(\sin\theta_i\cos\phi_i, \sin\theta_i\sin\phi_i, \cos\theta_i)$.

The important STT term $\vec{T}_{stt}=-a_I \vec{m}_i\times (\vec{m}_i\times \hat{s})$, $(i=1,2)$, taking the Slonczewski's form\cite{Slonczewski1} where $a_I=\frac{\hbar I P}{2\mu_0 e M_s^2 V}$ has a (normalized) magnetic field dimension. We omit the field-like STT term\cite{Zhang} because it is usually smaller than the Slonczewski's term. Furthermore, it has the magnetic field character whose role for the two-body system was already discussed previously\cite{Sunddi}. Here (among standard notation) $I$ and $P$ are the magnitude and spin polarization degree of the current, respectively. $\hat{s}$ is the unit vector along the current polarization direction (see Fig.~\ref{fig1}) which is characterized by two angles $\theta_s$ and $\phi_s$ in the spherical coordinates. Since $a_I$ is proportional to the current $I$ (assuming other parameters are constant), we will refer $a_I$ as to the switching current in the following description. For instance, $a_{I,c}$ stands for the critical switching current.

\section{Analytical results}

In the following we will concentrate on the synchronized motion mode of the
two dipoles in the current-induced switching. That is to say, both magnetization vectors
remain in parallel throughout the motion, $\theta_1=\theta_2=\theta$, and
$\phi_1=\phi_2=\phi$. Analytical results can be obtained in such motion
and the stability of the synchronization will be numerically verified later.
Thus, the two-body Stoner particles promise to play the role of an
information bit in computer technology. The modified LLG equations (\ref{LLG1})
in spherical coordinates are a system of coupled nonlinear equations and read
\begin{align}
\dot{\theta} +\alpha\sin\theta\dot{\phi}=&
a_I\frac{\partial \cos\psi_s}{\partial \theta}-3\eta\cos\psi_n\sin\theta_n\sin\phi,\nonumber\\
\alpha\dot{\theta}-\sin\theta\dot{\phi}=&
a_I\sin\theta_s\sin(\phi-\phi_s)-k\sin2\theta\nonumber\\
&+\frac{3\eta}{2}\frac{\partial \cos^2\psi_n}{\partial \theta}.\label{llg2a}
\end{align}
Here $\psi_{n/s}$ is the angle between $\vec{m}$ and $\hat{n}$/$\hat{s}$ such that $\cos\psi_{n/s}=\cos\theta\cos\theta_{n/s} +\sin\theta\sin\theta_{n/s}\cos(\phi-\phi_{n/s})$, (Note that we define $\phi_n=0$). Eq.~\eqref{llg2a} is the starting point for the numerical simulations.

In order to find the critical switching current $a_{I,c}$, one should find the fixed points (FPs) of the nonlinear equations~\eqref{llg2a} and then analyze their stability conditions\cite{fixedpoint}. Therefore, we first examine the FP condition for the two poles $\theta=0,\pi$. From Eq.~\eqref{llg2a}, we obtain
\begin{align}
&(3/2)\eta\sin 2\theta_n (\alpha \cos\phi\mp \sin\phi) \nonumber\\
&+a_I\sin\theta_s [\alpha\sin(\phi-\phi_s)
\pm \cos(\phi-\phi_s)]=0,
\end{align}
where $\mp$ or $\pm $ corresponds to $\theta=0$ and $\theta=\pi$, respectively. Due to the uncertainty of $\phi$ at poles, we conclude only $\sin2\theta_n=0$ and $\sin\theta_s=0$ such that the two poles become the possible FPs. Thus, we only need to focus on the simple cases of $\theta_s=0,\pi$ and $\theta_n=0,\pi, \pi/2$ (i.e. PARA and PERP configurations). In the following, we let $\theta_s=\pi$, which is the equivalent situation for $\theta_s=0$ by inverting the current flow direction.

Now let us first recover the usual case without DDI (i.e. $\eta=0$), where PERP and PARA configurations are indistinguishable. The motion of $\theta$ is decoupled to $\phi$,
\begin{equation}
\Gamma \dot{\theta}=a_I\sin\theta -\alpha k\sin2\theta,
\end{equation}
where we define $\Gamma\equiv1+\alpha^2$. The critical switching current $a_{I,c}$ can be obtained by analyzing the linear stability conditions at poles\cite{fixedpoint}. After some algebra one can easily find when $a_I > 2\alpha k$ the FP of $\theta=0$ becomes unstable, but $\theta=\pi$ is stable, which implies that the magnetization reversal switches on and $a_{I,c}$ reads
\begin{equation}
a_{I,c}^{0}=2\alpha k, \label{aic0}
\end{equation}
where the superscript index $0$ denotes the zero DDI case. We also like to point out that $2k$ is the usual Stoner-Wohlfarth limit\cite{Stoner} in the field reversal case, namely the barrier height between the two minima at poles in the energy landscape.

We now turn to the case with DDI ($\eta \neq 0$). Let us first write down the magnetostatic energy (without external field) under the synchronized motion mode:
\begin{equation}
E=-2k\cos^2\theta +\eta (1-3\cos^2 \psi_n),
\end{equation}
which immediately leads to the equilibrium (minimal) energy for the PARA and PERP configurations as
\begin{align}
E^{PARA}&=\eta - (2k+3\eta)\cos^2\theta, \nonumber\\
E^{PERP}&=-2\eta+(3\eta-2k)\cos^2\theta. \label{energy1}
\end{align}
Fig.~\ref{energy} shows that the essential differences in the energy landscapes for the PARA and PERP configurations. Figs.~\ref{energy}(a)-(c) show the energy variation in PARA configuration with the increase of the DDI strength $\eta$. The two edges of $\theta=0,\pi$ should be understood as the two poles in a sphere. In PARA cases, the energy diagrams have no essential geometrical shape change with variation of $\eta$. The two poles ($\theta=0,\pi$) are always the energy minima, and the in-between barrier (at $\theta=\pi/2$) height increases with $\eta$, expecting a higher critical switching current (or field) in the PARA case.

However, in the PERP configurations as shown in Figs.~\ref{energy}(d)-(f), the energy landscapes exhibit an essential geometrical change with the increase of $\eta$. Remarkably, from Eq.~\eqref{energy1}, a critical DDI strength $\eta_c=2k/3$ exists, which is the same value both in the current and field switching cases. Furthermore, $\eta_c$ is equivalent for the interparticle distance being $d_c=(3\mu_0M_s^2V/8\pi K)^{1/3}$, where $K=k\mu_0M_s^2$ is the standard anisotropy coefficient\cite{Sunddi}. At the critical distance, the energy diagram has two flat paths along $\phi=0,\phi$ [see Fig.~\ref{energy}(e)], which means any position of $\theta$ can be equilibriums. While on the both sides of the critical $\eta_c$ value, the system synchronized ground states (i.e. energy minimum) change. When the interparticle distance $d>d_c$ ($\eta<\eta_c$), the ground states are at two poles of $\theta=0, \pi$ [Fig.~\ref{energy}(d)] while when $d<d_c$ ($\eta>\eta_c$), the ground states are at another two positions, $\theta=\pi/2$ and $\phi=0,\pi$, along the hard $x$-axis.

\begin{figure}[htbp]
 \begin{center}
\includegraphics[width=6.5cm]{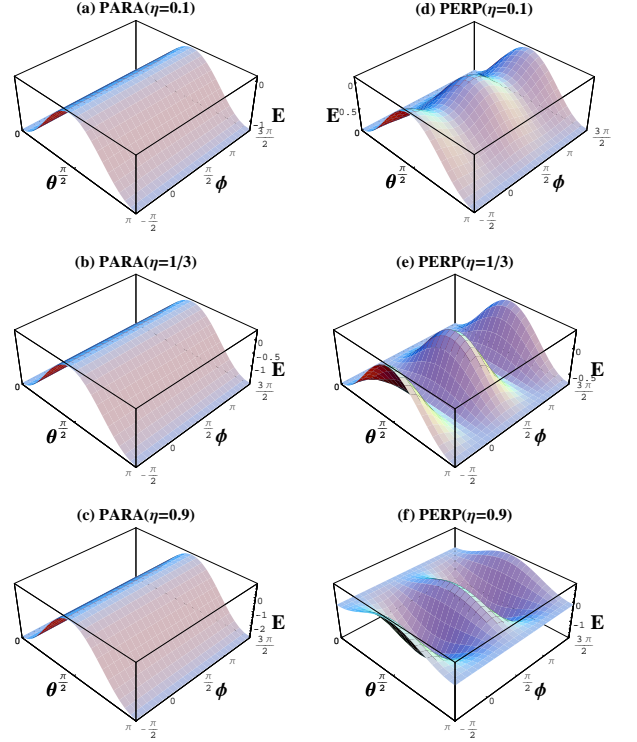}
 \end{center}
\caption{\label{energy} (color online.) The three-dimensional energy diagrams for the PARA and PERP configurations. (a)-(c): PARA with $\eta=0.1,1/3,0.9$; (d)-(f): PERP with $\eta=0.1,1/3,0.9$. The system parameter $k=0.5$. }
\end{figure}

For the PARA configuration, one can find the decoupled equation of $\theta$ to $\phi$,
\begin{equation}
\Gamma \dot{\theta}=a_I\sin\theta -\alpha (k+3\eta/2)\sin2\theta.
\end{equation}
With the same linear stability analysis as before, one can derive that when $a_{I} > \alpha (2k+3\eta)$, the FP of $\theta=0$ becomes unstable, but $\theta=\pi$ is still stable. Hence, the critical switching current in the PARA case reads,
\begin{equation}
a_{I,c}^{PARA}=\alpha (2k+3\eta). \label{aicpara}
\end{equation}
Thus, $a_{I,c}^{PARA}$ is always higher than $a_{I,c}^{0}$ for $\eta \neq 0$, as is expected that the barrier height between two energy minima is always increased with increasing $\eta$.

The PERP configuration is much more interesting and exhibits novel physical behavior due to the geometrical structure change in the energy diagrams as shown in Figs.~\ref{energy}(d)-(f), in which a novel zero-field switching mechanism was proposed\cite{Sunddi}. In PERP, the nonlinear Eq.~\eqref{llg2a} cannot be decoupled any more for $\theta$ and $\phi$,
\begin{align}
\Gamma \dot{\theta} &= a_I \sin\theta-\alpha k\sin2\theta\nonumber\\
& + (3/2)\eta(\alpha\sin2\theta\cos^2\phi-\sin\theta\sin2\phi),\\
\Gamma \dot{\phi}&=\alpha a_I+2k\cos\theta -3\eta\cos\theta\cos^2\phi\nonumber\\
&-(3/2)\alpha\eta\sin2\phi.\label{llgperp}
\end{align}
Again according to the linear stability analysis approach, the linearized matrices at the FPs of $\theta=0,\pi$ can be found as
$A|_{\theta=0,\pi}={\rm diag}(A_{11}^{\pm},A_{22}^{\pm})$ with
\begin{eqnarray}
A_{11}^{\pm} & = & \pm a_I-2\alpha k+3\eta
\left(\alpha\cos^2\phi\mp (\sin2\phi)/2\right)\\
A_{22}^{\pm} & = & 3\eta(\pm \sin2\phi-\alpha\cos2\phi)
 \label{matrix}
\end{eqnarray}
where diag denotes the diagonal matrix and the upper (lower) sign refers to  $\theta=0$ ($\theta=\pi$).
After some algebra, one finds that for $a_{I}> \alpha (2k-3\eta/2)$,
$\theta=0$ becomes unstable but $\theta=\pi$ is still stable, such that the
critical switching current reads,
\begin{equation}
a_{I,c}^{PERP,1}=\alpha (2k-3\eta/2), \quad (\eta<\eta_c). \label{aicperp1}
\end{equation}
The obtained $a_{I,c}^{PERP,1}$ can only be applicable for the regime of $\eta<\eta_c$ because the two-body system in the PERP case changes its energy equilibriums for $\eta>\eta_c$, where $\theta=\pi/2$ may become the FP, as shown in energy diagram Figs.~\ref{energy}(d)-(f). Thus, one requires to analyze the stability condition around the new FP for the $\eta>\eta_c$ regime, which can be obtained under the current case as:
\begin{equation}
\theta^{*}=\pi/2, \qquad \phi^{*} =\left(\sin^{-1}(2a_I/3\eta)\right)/2.
\end{equation}
Then, the linearized matrix at this FP is
\begin{equation}
 \label{matrix}
A|_{\theta^{*},\phi^{*}}=\begin{pmatrix}
2\alpha k-3\alpha
\eta \cos^2\phi^{*} & -3\eta\cos2\phi^{*}\\
3\eta\cos^2\phi^{*}-2k & -3\alpha \eta \cos2\phi^{*}
\end{pmatrix}.
\end{equation}
After some algebra, we can find that
when $4a_I^2+(4k-3\eta)^2>9\eta^2$, the FP becomes unstable, but the pole $\theta=\pi$ is still stable. Thus the critical switching current within the $\eta>\eta_c$ regime reads
\begin{equation}
\label{aicperp2}
a_{I,c}^{PERP,2}=\sqrt{2k(3\eta-2k)}, \quad (\eta>\eta_c).
\end{equation}

\begin{figure}[htbp]
 \begin{center}
\includegraphics[width=6.5cm]{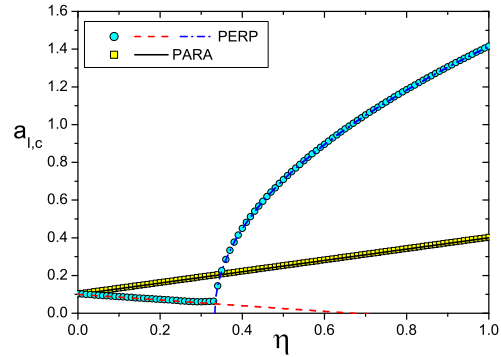}
 \end{center}
\caption{\label{swcur} (color online.) The (normalized) critical switching current $a_{I,c}$ versus the DDI strength $\eta$ for the PERP and PARA configurations. The lines show analytical results and the circles and squares show numerical results. The system parameters are $k=0.5$ and $\alpha=0.1$. }
\end{figure}

So far we obtained the analytical solutions of the critical switching current in both the PARA and PERP configurations, $a_{I,c}^{PARA}$, $a_{I,c}^{PERP,1}$ and $a_{I,c}^{PERP,2}$, which are plotted by the solid, dashed and dot-dashed lines in Fig.~\ref{swcur}, respectively. The system parameters $k=0.5$ and the damping $\alpha=0.1$. In principle, the minimum switching current in PERP should occur at the critical DDI value $\eta_c$,
\begin{equation}
a_{I,c}^{min}=\alpha k=a_{I,c}^{0}/2,
\end{equation}
which is obtained by using Eq.~\eqref{aicperp1} at $\eta_c$ and is a half of the value without DDI. However, as we will show later [see Fig.~\ref{self}(a)], there is a very small region around $\eta_c$ above the crossed two solutions Eq.~\eqref{aicperp1} and Eq.~\eqref{aicperp2}, where the magnetization dynamics is not static but shows self-precession. Thus, in practice, we numerically found that the minimal critical switching current $a_{I,c}^{min}$ for a static magnetization reversal can be lowered as about $70\%$ of the usual value $a_{I,c}^{0}$ without DDI, which will be more discussed later.

\section{Numerical results}

In order to complement and quantitatively support our previous discussions, we will now numerically verify our analytical results and investigate the reversal phenomena of two-body Stoner particles system in presence of magnetic dipolar interaction under a spin-polarized current. We solved the Eqs.~\eqref{llg2a} by implementing a fourth-order Runge-Kutta method. We consider the range of the DDI strength to be $0\leq \eta \leq 1$. In this case, $\eta=0$ corresponds to the limiting case of two particles infinitely apart from one another i.e. $d\rightarrow\infty$. Large $\eta$ may be realized by fabricating magnetic nanoparticles of ellipsoidal shapes allowing for a closer proximity. Throughout the numerical calculations, we use the damping parameter of $\alpha=0.1$. In Fig.~\ref{swcur}, we first compare the numerical simulations of the critical switching current $a_{I,c}$ in both PERP and PARA configurations (circles and squares) and show an excellent agreement to our analytical solutions: Eq.~\eqref{aicpara}, Eq.~\eqref{aicperp1} and Eq.~\eqref{aicperp2}. The critical switching current in PARA is always higher than the value without DDI (at $\eta=0$), while in PERP it bifurcates at the critical DDI value (at $\eta=1/3$) with a square-root dependence behavior.

\subsection{Hysteresis loop}

\begin{figure}[htbp]
 \begin{center}
\includegraphics[width=4.cm,height=3.5cm]{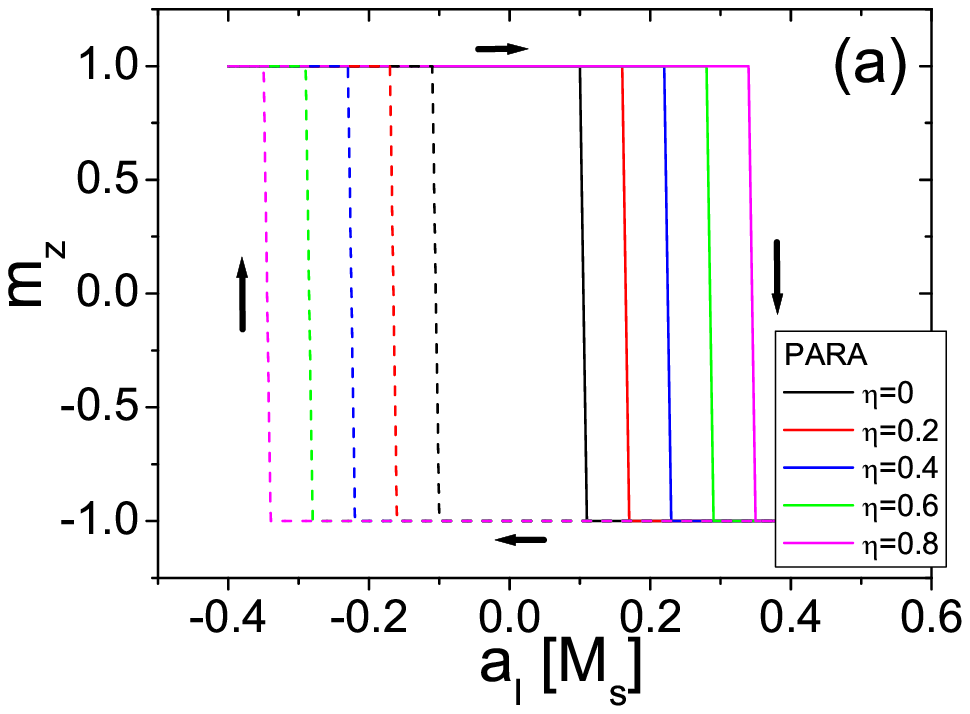}
\includegraphics[width=4.cm,height=3.5cm]{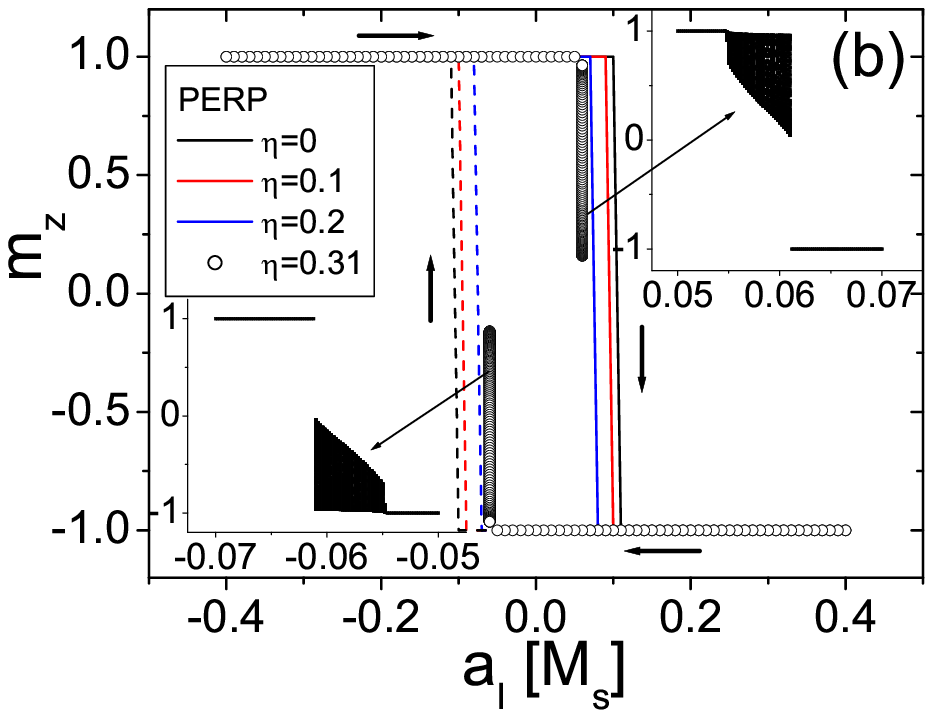}
\includegraphics[width=4.cm,height=3.5cm]{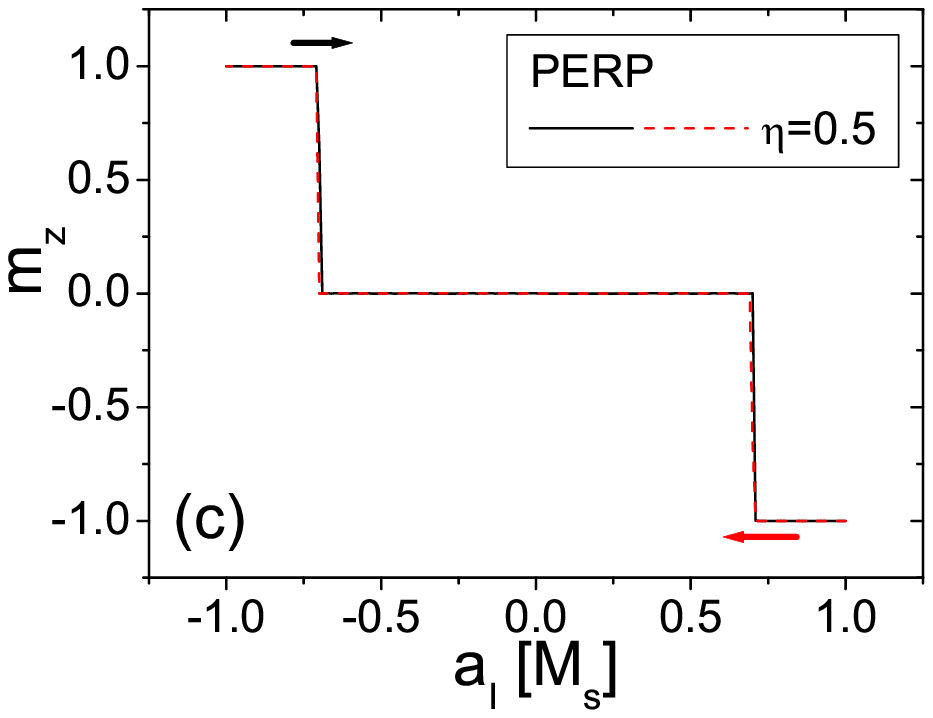}
\includegraphics[width=4.cm,height=3.5cm]{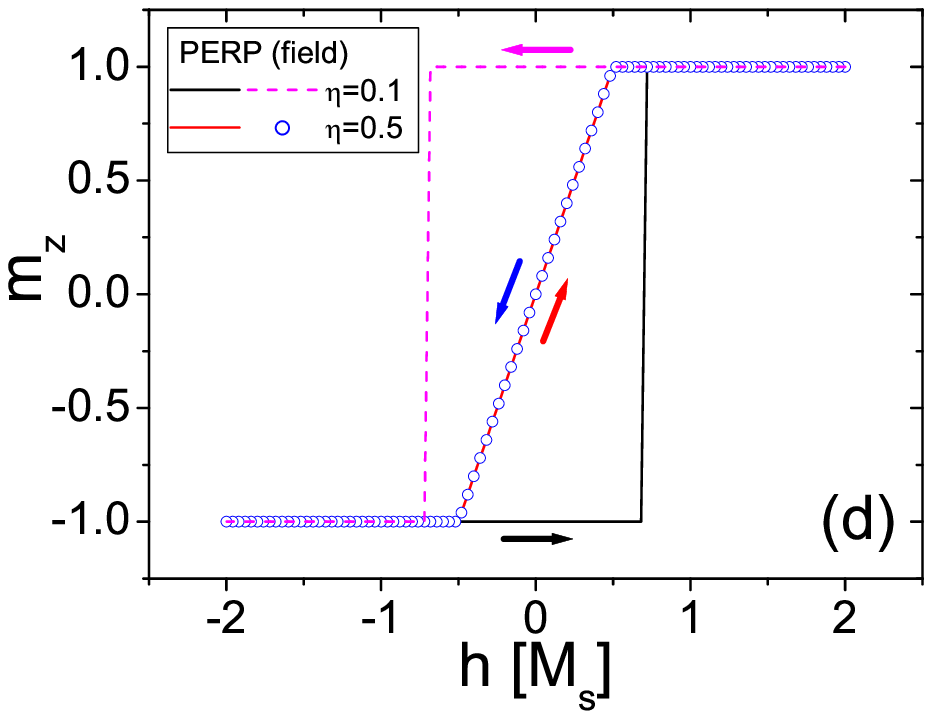}
 \end{center}
\caption{\label{hystere} (color online.) Hysteresis loops. (a) PARA configuration. (b) PERP configuration when $\eta<\eta_c=1/3$. The insets show the enlargement of the magnetic self-precession regimes, where $m_z$ is proportional to the oscillation amplitude. (c) PERP configuration when $\eta>\eta_c=1/3$. (d) PERP configuration in the field switching case. Other parameters are the same as Fig.~\ref{swcur}. }
\end{figure}

We then, with the numerical simulations, study the magnetic hysteresis loops under the current-induced synchronized switching. Fig.~\ref{hystere}(a) first shows the hysteresis loops at different DDI strengthes for the PARA configuration. The increase of $\eta$ will enlarge the loop sizes. Fig.~\ref{hystere}(b) shows the hysteresis loops for the PERP configuration for $\eta<\eta_c$. The increase of $\eta$ will reduce the loop sizes, in contrast with the PARA case. In particular, when $\eta$ approaches closely to $\eta_c$, magnetic self-precession phenomenon in current switching can be observed, which will be discussed in detail in next subsection. For example, there are very narrow regimes at $\eta=0.31$, which is enlarged in the insets of panel (b), showing the $m_z$ oscillation amplitudes. Fig.~\ref{hystere}(c) shows the hysteresis behavior for the PERP configuration for $\eta>\eta_c$. It is interesting to note that there is no any loops but a reversing tri-state occurring. At last, as a comparison, we plot panel (d) to show the hysteresis behaviors in the field switching case for the PERP configuration, where $h$ is the normalized external field by the unit of magnetization saturation $M_s$. When $\eta<\eta_c$ a clear hysteresis loop occurs while when $\eta>\eta_c$ there is neither hysteresis loops but the magnetization $m_z$ is linearly stable with $h$, different from the tri-state phenomenon in the current driven case.

\subsection{Self-precession}

\begin{figure}[htbp]
 \begin{center}
\includegraphics[width=4.cm, height=3.5cm]{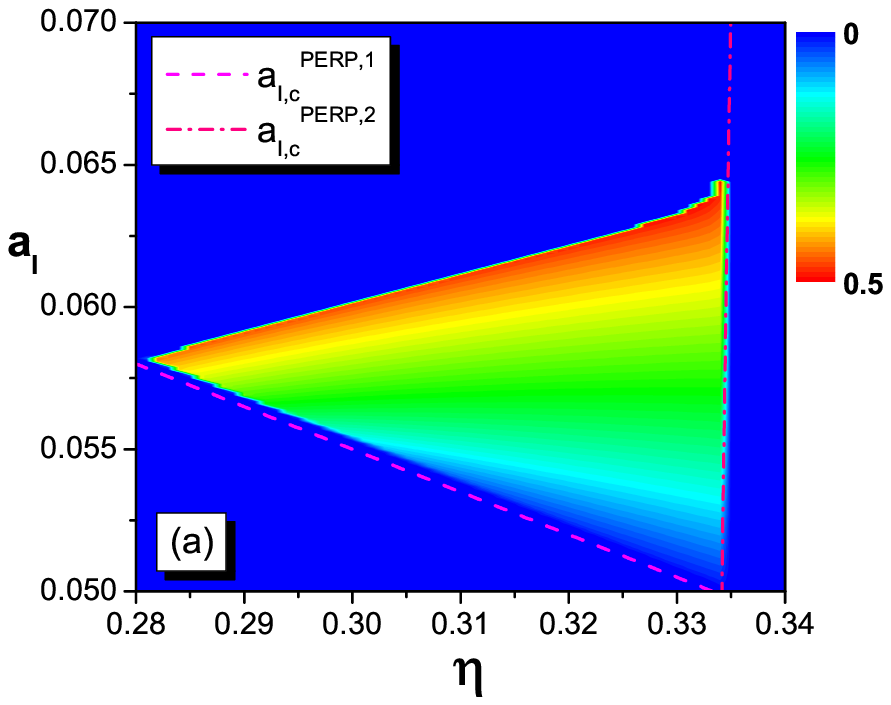}
\includegraphics[width=4.5cm, height=3.5cm]{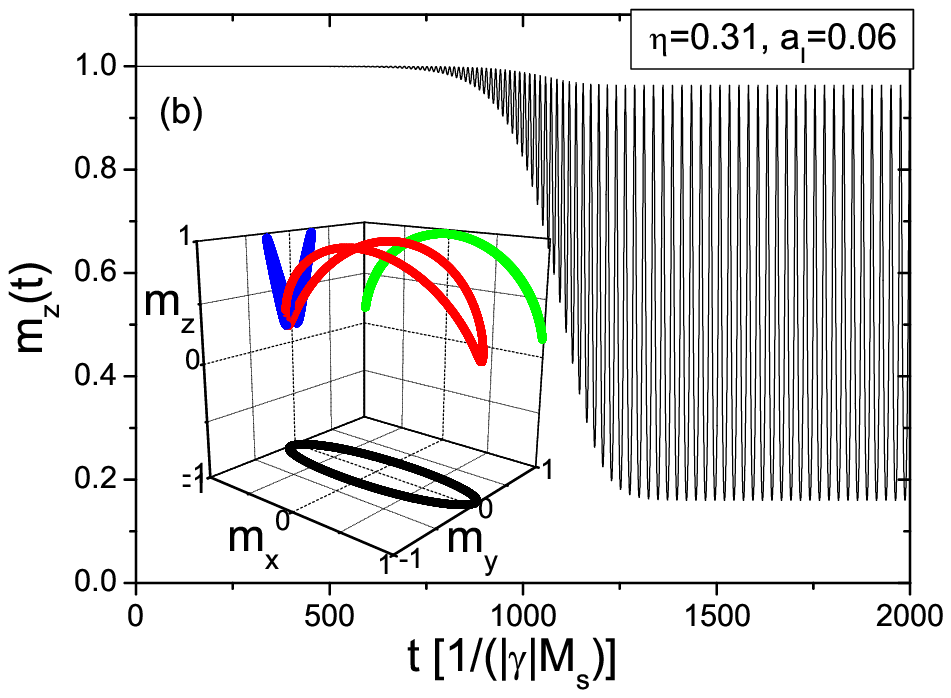}
 \end{center}
\caption{\label{self} (color online.) (a) The phase diagram in the PERP configuration of the magnetization self-precession in the parameter $\eta-a_{I}$ plane, where the color is proportional to the self-oscillation amplitude and the two lines show the analytical solutions of the critical switching current. (b) The self-oscillation curve of $m_z(t)$ at $\eta=0.31$ and $a_{I}=0.06$. The inset shows the spacial trajectories of the magnetization vector and its projections. Other parameters are the same as Fig.~\ref{swcur}.}
\end{figure}

With the numerical simulations, we observed in the PERP configuration, there exists a small region above the crossed two solutions Eq.~\eqref{aicperp1} and Eq.~\eqref{aicperp2} around the critical DDI value $\eta_c$. The magnetization vectors in this region perform the STT induced self-precession (or self-oscillation) phenomenon\cite{Slonczewski1,fixedpoint}. In Fig.~\ref{self}(a), we plot the self-precession phase diagram in the parameter $\eta-a_{I}$ space. The plotted color is proportional to the magnetic self-oscillation amplitude for $m_z(t)$. The self-precession occupies only a very small partition in the parameter space for the current-induced synchronized dynamics of two-body systems, which is consistent with the previous studies for single biaxial magnetic nanoparticle without DDI\cite{fixedpoint}.
Fig.~\ref{self}(b) shows an example of the self-oscillation of $m_z(t)$ in time domain at $\eta=0.31$ and $a_{I}=0.06$. The inset shows the spacial trajectories of the magnetization vector and its projections in $m_x-m_y-m_z$ coordinates. We also like to point out that the practical critical switching current $a_{I,c}^{min}$ at $\eta=\eta_c$ cannot be lowered to a half of that without DDI due to the self-precession phenomenon.

\subsection{Reversal time}

\begin{figure}[htbp]
 \begin{center}
\includegraphics[width=4.cm, height=3.5cm]{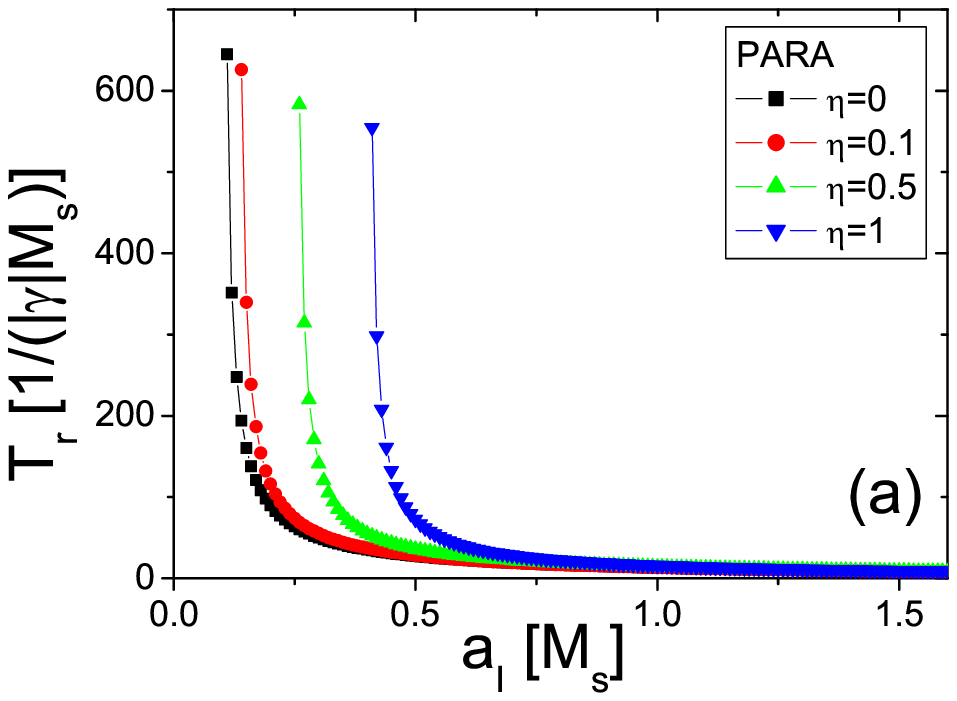}
\includegraphics[width=4.cm, height=3.5cm]{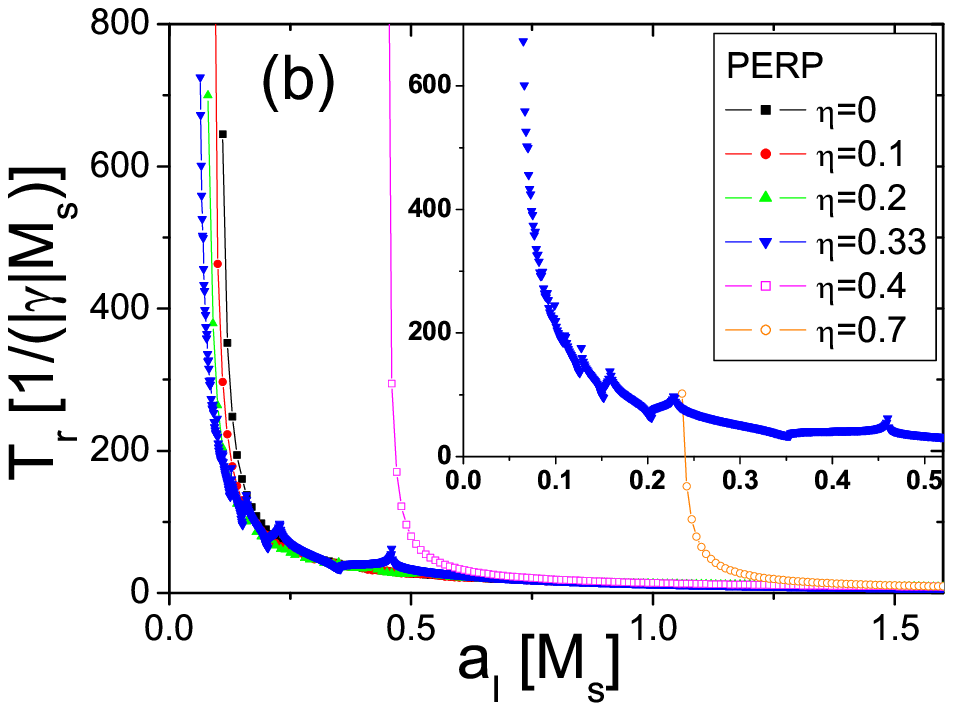}
 \end{center}
\caption{\label{time} (color online.) Reversal time $T_r$ versus the current magnitude $a_I$. (a) PARA configuration: different DDI strength $\eta =0, 0.1, 0.5, 1$; (b) PERP configuration: $\eta=0, 0.1, 0.2, 0.33, 0.4, 0.7$. The inset is an enlargement for $\eta=0.33$ where non-smooth $T_r$ is observed. The connected lines are guides to eyes. Other parameters are the same as Fig.~\ref{swcur}.}
\end{figure}

The numerical results for the reversal time $T_r$ versus the switching spin-polarized current magnitude $a_I$ at different DDI strength for both PARA and PERP configurations are shown in Fig.~\ref{time}(a) and (b). The reversal time is numerically determined from $m_{z}=+1$ to $m_{z}=-1$ for either particle, which depends on the initial directions of the magnetization vectors due to the metastability of the poles\cite{Sunddi}. Thus, in practice, we deviate the initial angles $\theta_1(0)=\theta_2(0)=0.001$ in order to find a finite reversal time. The same initial values for either particle is of importance for the stability of the synchronized motion mode, as we will discussed in next subsection. Fig.~\ref{time}(a) and (b) show that, in both the PARA and PERP cases, the reversal time curves with current have similar trends, approximately inversely proportional to the current magnitude $a_I$ for large currents. In the PERP configuration, the $T_r$ behavior is different from that in the field switching case when the DDI parameter approaches its critical value. A substantially shorter reversal time can be found around zero field regime in the field case\cite{Sunddi} when $\eta\approx \eta_c$. In contrast, the reversal time curve is non-smooth in the current case when $\eta$ is around its critical value, as shown in the inset of Fig.~\ref{time}(b).

\subsection{Stability phase diagram}

As an important issue, we now examine the stability of the synchronized magnetization dynamics scenario in the PERP configuration. We examine the stability from two aspects. One is that the difference in initial angles of the two magnetization vectors may result in unstablity for the synchronization. The other is the tiny difference in two particles' material parameters may spoil the synchronization.

\begin{figure}[htbp]
 \begin{center}
\includegraphics[width=4.cm, height=3.5cm]{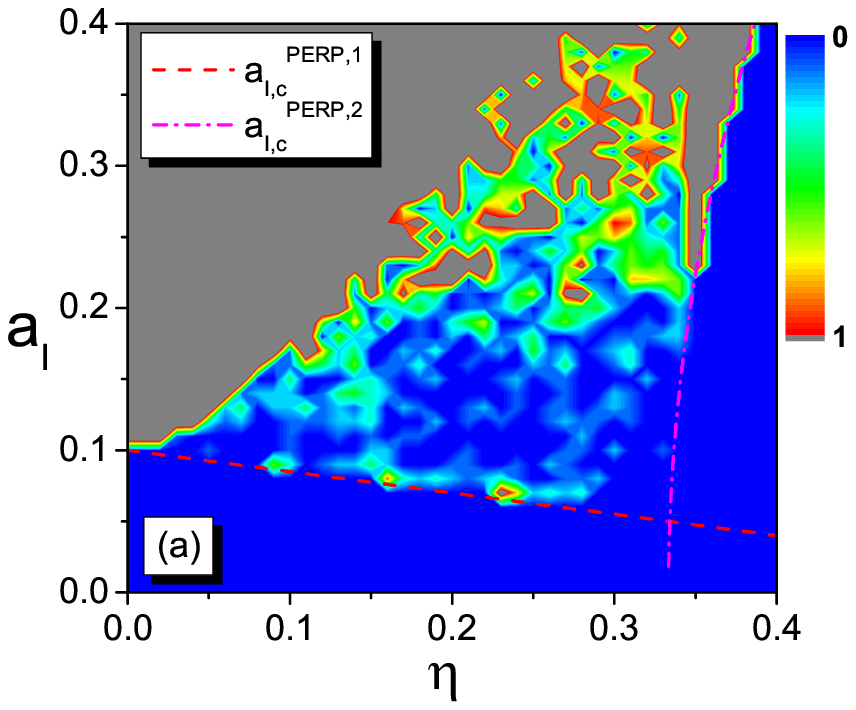}
\includegraphics[width=4.cm, height=3.5cm]{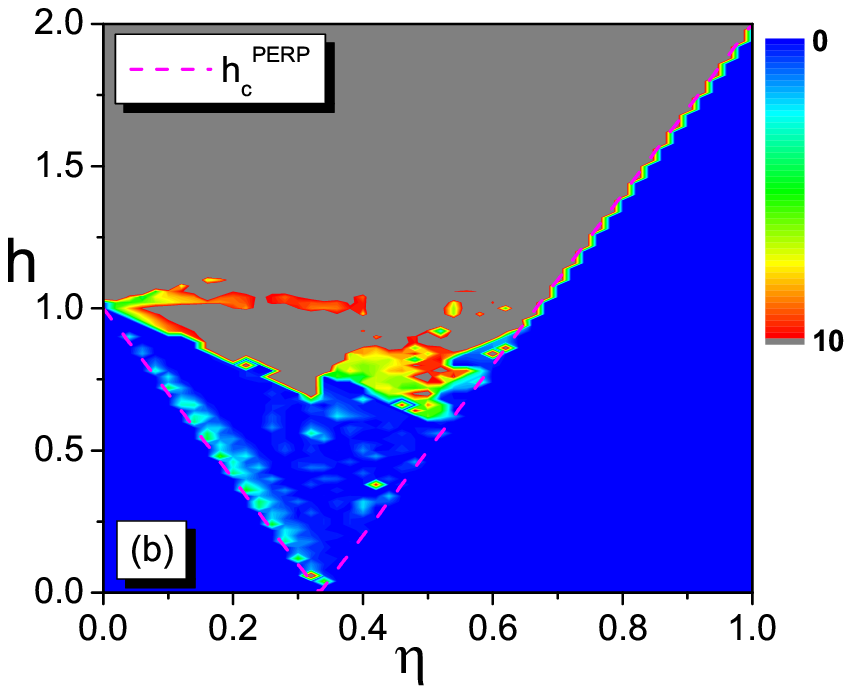}
\includegraphics[width=4.cm, height=3.5cm]{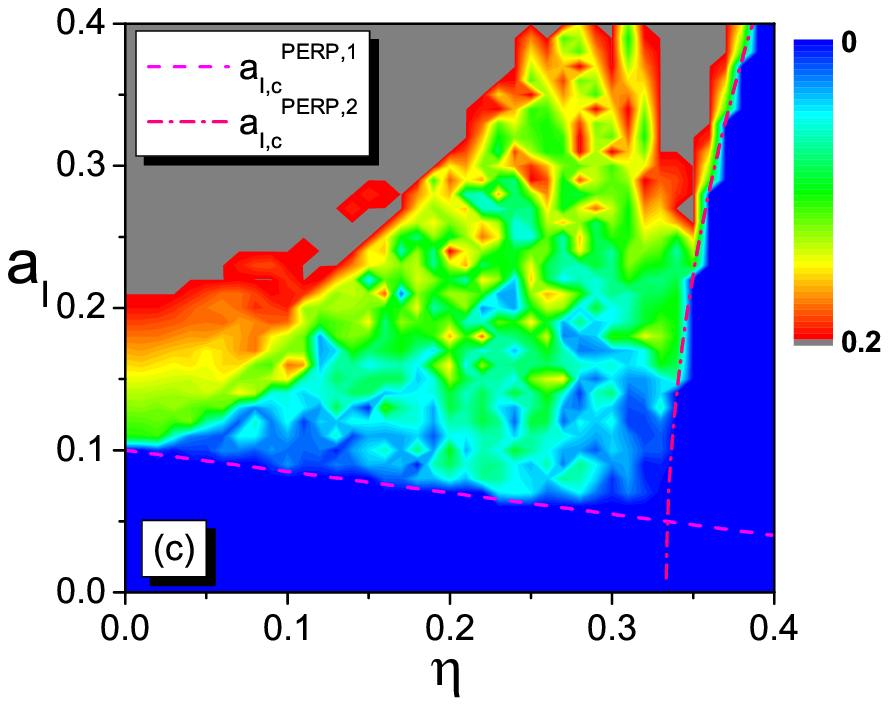}
\includegraphics[width=4.cm, height=3.5cm]{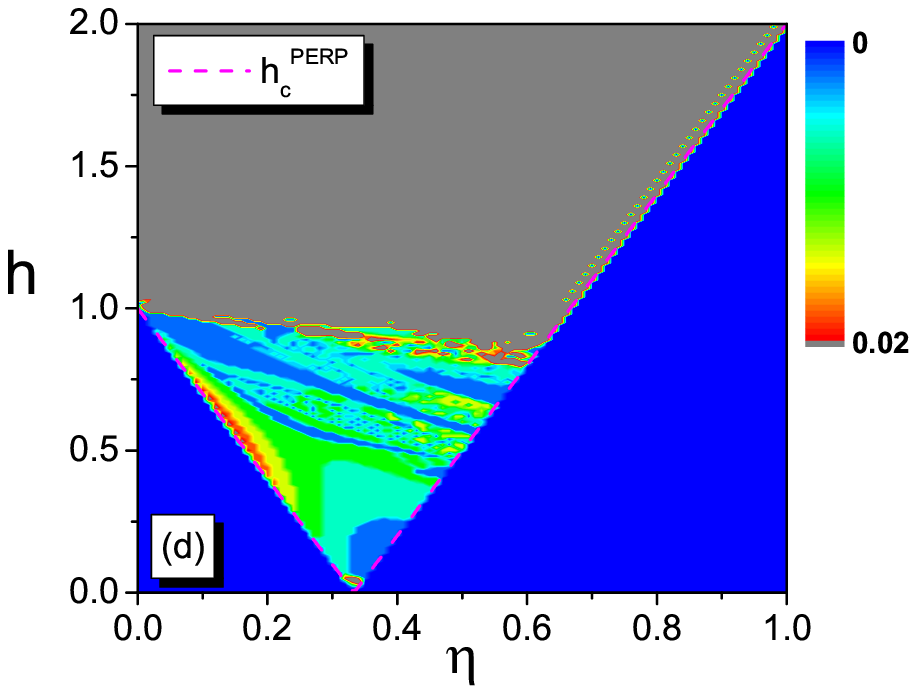}
 \end{center}
\caption{\label{phase} (color online.) Synchronization stability phase diagrams in the PERP configuration for the initial direction deviation $\delta$ of the magnetization vectors in current case (a); in field case (b). The color difference is proportional to maximal stability $\delta_m$ in unit of degrees where the gray color denotes $\delta>1^\circ$ in (a) and $\delta>10^\circ$ in (b). Synchronization stability phase diagrams for different experienced $\delta a_I$ in current case (c); $\delta h$ in field case (d). The color difference is proportional to the variation width $\delta a_I^{w}$ or $\delta h^{w}$ where the gray color denotes $\delta a_I^{w}>0.2$ and $\delta h^{w}>0.02$. Other parameters are the same as Fig.~\ref{swcur}.}
\end{figure}

\begin{figure}[htbp]
 \begin{center}
\includegraphics[width=6.5cm]{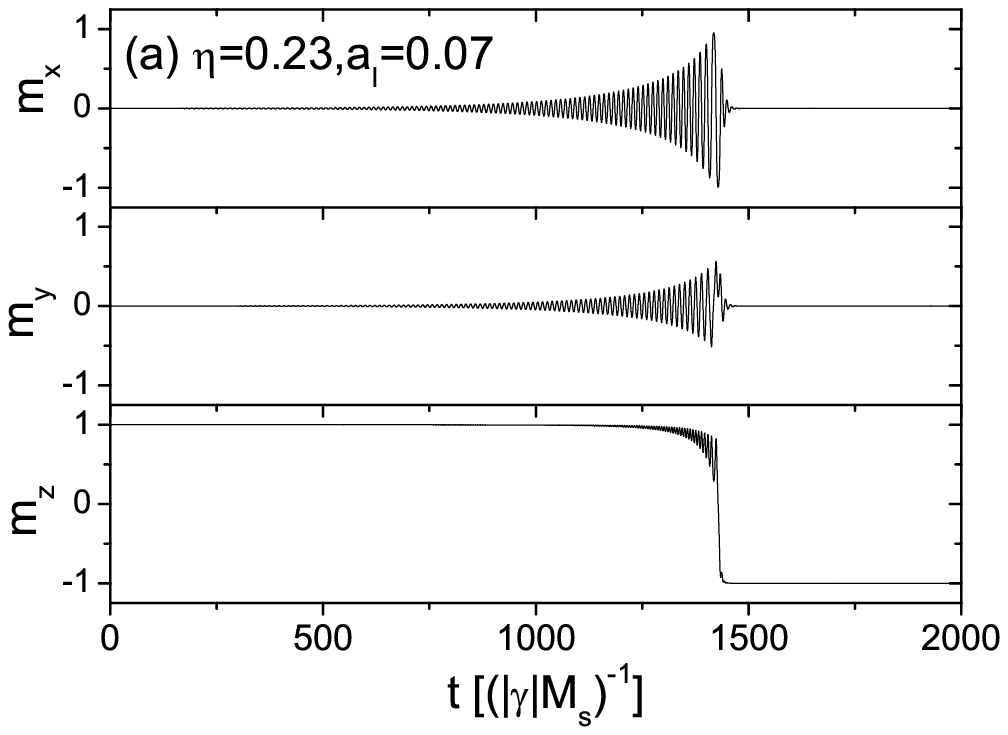}
\includegraphics[width=4.cm, height=3.cm]{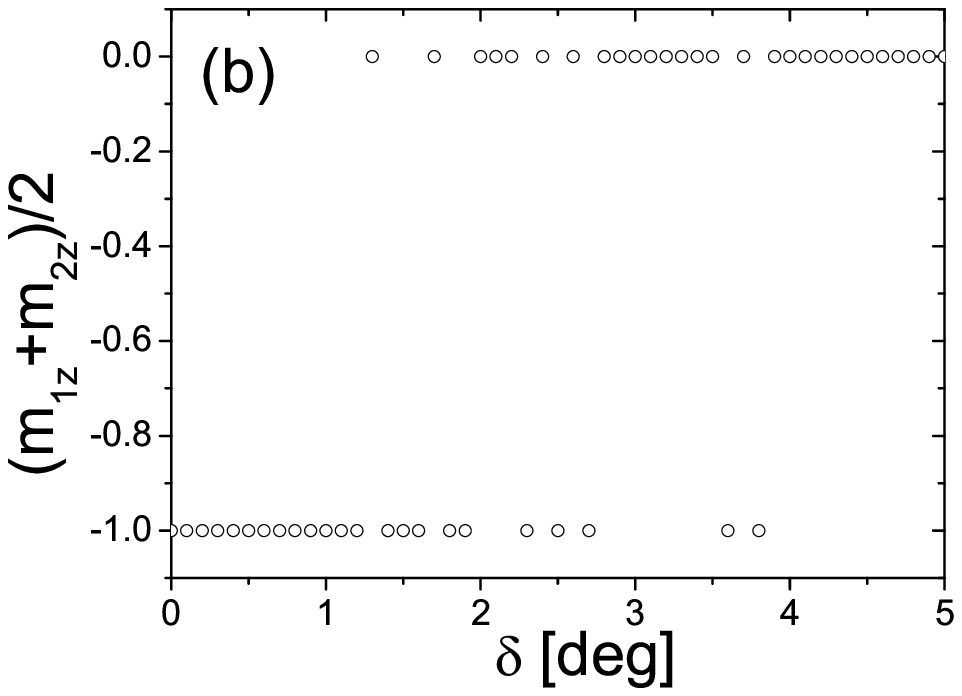}
\includegraphics[width=4.cm, height=3.cm]{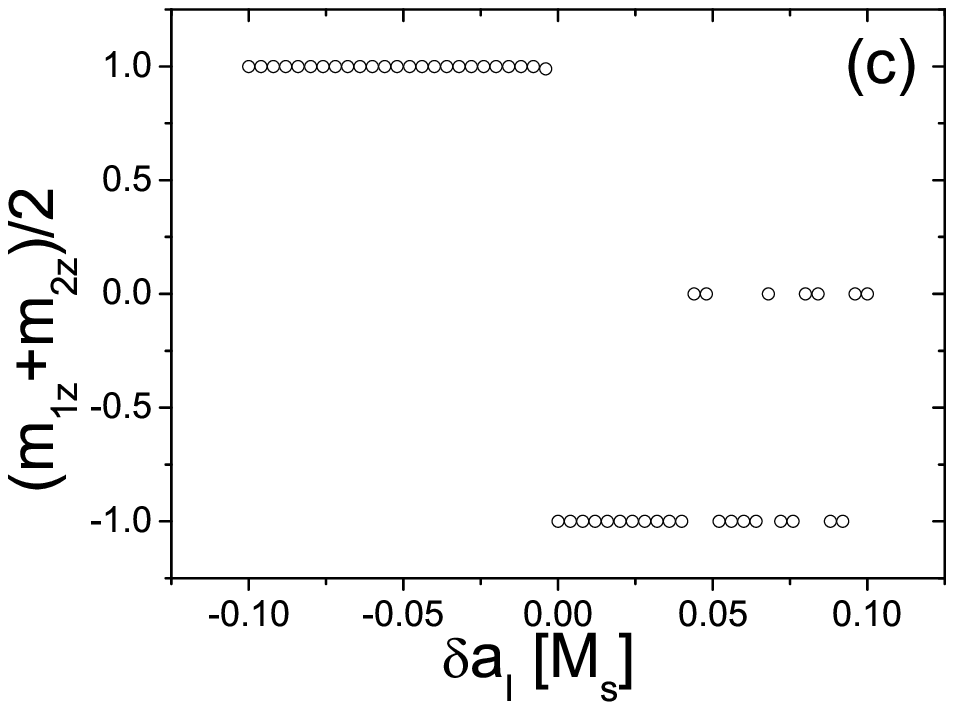}
 \end{center}
\caption{\label{mzt} (a) The magnetization vector components versus time at $\eta=0.23$ and $a_I=0.07$. (b) The average stable magnetization $(m_{1z}+m_{2z})/2$ versus the initial angle deviation $\delta$, where $\delta_m\approx 1.2^{\circ}$ is observed. (c) $(m_{1z}+m_{2z})/2$ versus the current deviation $\delta a_I$, where $\delta a_I^{w}\approx 0.04$ is found. Other parameters are the same as Fig.~\ref{swcur}. }
\end{figure}

Let us first numerically examine the average stable magnetization $(m_{1z}+m_{2z})/2$ (the value $-1$ means stable while $0$ unstable) with the dependence of the deviation angle $\delta$ between the initial directions of the two magnetization vectors. In detail, we fixed the initial direction of one particle to be along the $z$-axis i.e. $\theta_1=0.001$ and changed the other one as $\theta_2=0.001+\delta$ where $0.001$ is for eliminating the singularity in poles. We can plot the stability phase diagram in the $a_I-\eta$ space, as shown in Fig.~\ref{phase}(a) for the current case, and plot Fig.~\ref{phase}(b) for the field switching case as a comparison in the $h-\eta$ space. The color difference is proportional to maximal $\delta_m$ in unit of degrees for guaranteeing the stability where the gray color denotes $\delta_m>1^\circ$ in (a) and $\delta_m>10^\circ$ in (b). That is to say, when $\delta>\delta_m$, the final average magnetization is found to be $0$, implying the synchronized dynamics is destroyed. One can observe two points from Fig.~\ref{phase}(a) and (b): (1) In both current and field switching cases, the important stability regimes are located just above the critical switching current or field lines shown in dashed or dot-dashed lines in corresponding figures, especially for $\eta<\eta_c$. ``Islands" appear for such regions. (2) It has wider initial deviation angles limit $\delta_m$ for the synchronization stability in the field case ($<10^{\circ}$) than in the current case ($<1^{\circ}$). We also plot an example shown in Fig.~\ref{mzt} in the case of $\eta=0.23$ and $a_I=0.07$, which is $70\%$ of the usual switching current $a_{I,c}^0=0.1$ without DDI. Fig.~\ref{mzt}(a) shows the time evolution of all the components of the magnetization vector. Fig.~\ref{mzt}(b) shows the maximal initial deviation angle $\delta_m \approx 1.2^{\circ}$ for the synchronization stability. Consequently, we remark that the practical critical switching current $a_{I,c}^{min}$ for a static magnetization reversal can be lowered as about $70\%$ of the usual value $a_{I,c}^{0}$.

On the other hand, for two particles with slightly different, for instance, anisotropy $K_i$, volume $V_i$, saturation magnetization $M_{s,i}$ ($i=1,2$), as discussed before\cite{Sunddi}, the critical value of DDI becomes $\eta_c=(k_1+k_2)/3=\frac{2(K_1V_1+K_2V_2)}{3\mu_0M_{s1}M_{s2}(V_1+V_2)}$ and the effective field experienced by each particle due to the external field or the STT here will be slightly different. Hence, we also numerically examine the synchronization stability due to such difference. In detail we fixed the current torque on one particle as $a_{1,I}$ and changed the other one as $a_{2,I}=a_{1,I}+\delta a_I$ where $\delta a_I$ denotes a small deviation for different currents. The field case is the same except introducing a small deviation $\delta h=h_2-h_1$ for different external fields. We thus plot the stability phase diagrams in the $a_I-\eta$ space, as shown in Fig.~\ref{phase}(c) for the current case, and Fig.~\ref{phase}(d) for the field case as a comparison in the $h-\eta$ space. The color difference is now proportional to the variation width of $\delta a_I^{w}$ and $\delta h^{w}$, in which the final average magnetization is found to be $-1$ for stability. The gray color denotes $\delta a_I^{w}>0.2 (M_s)$ in (c) and $\delta h^{w} >0.02 (M_s)$ in (d). From the figures, one can observe the stability ``islands" structures above the critical switching current or field lines become smeary now, and the current switching case has a wider stability window due to the parameter differences of the two particles than that in the field case. Fig.~\ref{mzt}(c) shows an example for current difference on two particles at the case of $\eta=0.23$ and $a_I=0.07$, where the width $\delta a_I^{w} \approx 0.04$ for the average stable magnetization equaling $-1$ to sustain the synchronized motion mode.

\section{Discussion and conclusion}

We first like to compare our results with a concrete magnetic material such as cobalt (Co) particles. The standard data is $M_s=1400$kA/m, uniaxial strength $K=10^5$J/m$^3$, and $\alpha=0.1$\cite{Back}. Thus $k=K/(\mu_0M_s^2)= 0.04$ such that the critical DDI strength $\eta_c=0.027$. If we consider two spherical particles with the radius $r$ so that the DDI parameter $\eta=r^3/(3d^3)$. Thus the critical DDI is reached at $d_c=2.3r$. Assuming $r=10$nm and spin polarization $P=0.4$, the critical switching ``effective field" without DDI is $a_{I,c}^0=(2\alpha k)M_s=140$Oe, which translates the critical switching current density to be about $j_0=3\times 10^8 $A/cm$^2$. In our model with DDI we argue that the critical switching current can be lowed to be $70\%$ of the original value, i.e. about $2\times 10^8 $A/cm$^2$ when the two particles are engineered to be located near the critical distance $d_c=23$nm. Also in the case of Co, the unit of time is $(|\gamma|M_s)^{-1}\ =3.23$ps. From Fig.~\ref{time}(b), the reversal time is infinitely long at the critical switching current point. Increasing the switching current will drop the reversal time inversely. More importantly, the key issue for the technological aim in our two-body Stoner particles system is to maintain the synchronized motion mode against large deviations from such as initial conditions and/or material parameters. The stability region from the initial angle deviations in the current case (about $1^\circ$) is much smaller than that in the field case (about $10^\circ$), thus how to enhance the synchronization stability might be most interesting for future studies. Moreover whether there exists a zero switching current in contrast with the zero-field switching case with the aid of DDI is also an interesting issue.

In conclusion, we have investigated the magnetization reversal of two-body
uniaxial Stoner particles in a stable synchronized motion mode, by
injecting a spin-polarized current through a spin-valve like structure. In
presence of magnetic dipolar interaction, the critical switching current for
reversing the two dipoles is analytically obtained and numerically verified
in two typical geometric PERP and PARA configurations. In the interesting
PERP configuration, the critical switching current bifurcates at a critical
DDI strength with a square-root behavior, near where it can be lowered to
about $70\%$ of the usual value without DDI. Moreover, we also numerically
investigate the current-induced magnetization hysteresis loops, magnetic
self-precession phenomenon, reversal time and the synchronization
stability phase diagram in the two-body system, which shows interesting
predictions and is expected to be useful for future device applications.

Z.Z.S. thanks the Alexander von
Humboldt Foundation (Germany) for a grant.
This work has been supported by Deutsche Forschungsgemeinschaft via SFB 689.

\end{document}